\documentclass[showpacs,preprintnumbers,amsmath,amssymb]{revtex4}

\usepackage{graphicx}
\usepackage{dcolumn}
\usepackage{bm}
\usepackage{url}
\usepackage{epsfig}
\usepackage{multirow}
\usepackage{tensor}
\usepackage{empheq}
\usepackage{amsmath}
\usepackage{color}
\definecolor{myblue}{rgb}{.8, .8, 1}
\usepackage{makecell,diagbox}

\def\be{\begin{equation}}
\def\ee{\end{equation}}
\def\ba{\begin{eqnarray}}
\def\ea{\end{eqnarray}}

\newcommand{\fr}[2]{\frac{#1}{#2}}

\def\D{\rm{D}}

\def\ga{\mathrel{\raise.3ex\hbox{$>$\kern-.75em\lower1ex\hbox{$\sim$}}}}
\def\la{\mathrel{\raise.3ex\hbox{$<$\kern-.75em\lower1ex\hbox{$\sim$}}}}

\begin{document}


\leftline{KIAS-P15036}

\title{New Method for Probing Dark Energy using the Rees-Sciama Effect}


\author{Seokcheon Lee}

\affiliation{School of Physics, Korea Institute for Advanced Study, Heogiro 85, Seoul 130-722, Korea}



\begin{abstract}
The integrated Sachs-Wolfe (ISW) effect provides us the information of the time evolution of gravitational potential. The cross-correlation between the cosmic microwave background (CMB) and the large scale structure (LSS) is known as a promising way to extract the ISW effect. Compared to CMB, the matter fluctuation can grow non-linearly and this is represented in the gravitational potential. Compared to the linear ISW effect, this non-linear ISW effect known as the Rees-Sciama (RS) effect shows the unique behavior by changing the anti-correlated cross correlation between the CMB and the mass tracer into the positively correlated cross correlation. We show that the dependence of this flipping scale on dark energy models and it might be used as a new method to investigate dark energy models.   

\end{abstract}



\maketitle

\setcounter{equation}{0}
The cosmic microwave background (CMB) emanating from the last scattering surface interacts with the intervening large scale structure (LSS) on its way to the present observer. This causes the secondary anisotropy of the fluctuation of the CMB temperature, which is called the integrated Sachs-Wolfe (ISW) effect \cite{ISW}. If one include the late time evolution of the LSS, then one also needs to consider the full non-linear ISW effect known as the Rees-Sciama (RS) effect \cite{RS}.

The temperature fluctuation ($\Delta T / T \equiv \Theta$) due to the ISW effect in the flat Universe is given by the line of sight integral of the change in the gravitational potential to the last scattering surface,
\be \Theta^{\rm{ISW}}(\hat{n}) = \sum_{l,m} a_{lm}^{\rm{ISW}} Y_{lm}(\theta,\phi) = \fr{1}{c} \int_{\eta_{\rm{ls}}}^{\eta_{0}} d \eta e^{-\tau} (\Phi' + \Psi') [\eta, \hat{n}(\eta_0-\eta)]   \simeq \fr{2}{c} \int_{\eta_{\rm{ls}}}^{\eta_{0}} d \eta \Phi' [\eta, \hat{n}(\eta_0-\eta)]\label{ThetaISW} \, , \ee
where $\eta$ is the conformal time, $\eta_0$ being today, $\eta_{\rm{ls}}$ being recombination, $\tau$ is the optical depth, primes mean the derivatives with respect to the conformal time, $\Phi$ is the Newtonian potential, and $\Psi$ is the spatial curvature perturbation, respectively. Because we restrict our consideration to the perfect fluid with the general relativity and also assume the instantaneous reionization in this {\it letter}, we ignore the anisotropic stress tensor and the optical depth in the last equality of the above equation (\ref{ThetaISW}). 

The ISW effect from dark energy can be detected with the cross-correlation function between the CMB anisotropies and the projected galaxy density \cite{9510072, 9506048}. This is used for the source of observed hot and cold spots in the CMB temperature around known galaxy clusters and cosmic voids \cite{08053695, 150201595}.
The Newtonian potential $\Phi$ is related with the matter density fluctuation $\delta$ through the Poisson equation 
\be \Phi(\vec{k},\eta) = -\fr{3}{2} \fr{\Omega_{{\rm m} 0}}{a(\eta)} \Biggl(\fr{H_0}{c k} \Biggr)^2 \Biggl[ \delta (\vec{k},\eta) + 3 \fr{a H}{k^2} \theta(\vec{k}, \eta) \Biggr] \simeq  -\fr{3}{2} \fr{\Omega_{{\rm m} 0}}{a(\eta)} \Biggl(\fr{H_0}{c k} \Biggr)^2 \delta (\vec{k},\eta)\label{Phi} \, , \ee
where $\Omega_{{\rm m}0}$ is the present matter energy contrast, $H_0$ is the present value of Hubble parameter, $\delta$ is the matter density fluctuation, and $\theta$ is the divergence of the peculiar velocity, respectively. We use the Newtonian limit where the scale of interest is much smaller than the horizon, {\it i.e.} $k \gg aH$, in the  approximation of the above equation (\ref{Phi}) to obtain the last equality. Thus, $\Phi$ is negative (positive) for the over (under) density region.

The cross-correlation between the ISW and the density tracer is useful to isolate the ISW effect from the CMB. Then the cross-correlation between the ISW and any tracer of the density field can be some functions of the cross-power spectrum of $\Phi$ and $\Phi'$. From the Poisson equation, one obtains
\be \Phi'(\vec{k},\eta) = -\fr{3}{2} \fr{\Omega_{{\rm m} 0}}{a(\eta)} \Biggl(\fr{H_0}{c k} \Biggr)^2 \Bigl( \delta' - {\cal H} \delta \Bigr) \simeq -\fr{3}{2} \fr{\Omega_{{\rm m} 0}}{a(\eta)} \Biggl(\fr{H_0}{c k} \Biggr)^2 ( f -1 ) {\cal H} \delta \label{dotPhi} \, , \ee
where we use the linear approximation in the second equality, ${\cal H} \equiv \fr{1}{a} \fr{\partial a}{\partial \eta}$,  and $f \equiv \fr{d \ln \delta}{d \ln a}$ is the growth rate. Thus, $\Phi'$ is always positive (negative) for the over (under) dense region in the linear regime. However, $\delta'$ can be greater than ${\cal H} \delta$ in the non-linear regime and it causes the sign change of $\Phi'$. If one considers the cross-correlation between $\Phi$ and $\Phi'$, then it is always anti-correlated in the linear regime, but it can be positively correlated in the non-linear regime. The typical scale for this flipping of the cross-correlation from the anti-correlation to the positive correlation depends on dark energy model, and thus it can be used for the investigation of dark energy models. 

From Eq.(\ref{Phi}) and Eq.(\ref{dotPhi}), one obtains the cross-correlation power spectrum between the ISW and any tracer of the density field which is some functions of the cross power spectrum of $\Phi$ and $\Phi'$
\be P_{\Phi \Phi'} (k,\eta) \equiv \Bigl \langle \Phi(\vec{k},\eta) \Phi'(\vec{k},\eta)  \Bigr \rangle = \fr{9}{4} \fr{\Omega_{{\rm m} 0}^2}{a(\eta)^2} \Biggl(\fr{H_0}{c k} \Biggr)^4 \Bigl( P_{\delta \delta'}(k,\eta) - {\cal H} P_{\delta \delta}(k,\eta) \Bigr) \label{PPhidotPhi} \, . \ee
In order to consider the non-linear effect of the matter density fluctuation, one can use N-body simulation \cite{08094488, 10030974, 13016136}, a halo model \cite{0206508}, or the higher order perturbation theory \cite{08094488, 14045102}. We consider the standard perturbation theory to include the non-linear effect in the cross-correlation power spectrum. The equations of motion of the matter density fluctuation $\hat{\delta}$ and the divergence of the peculiar velocity $\hat{\theta}$ in the Fourier space are given by
\ba \fr{\partial \hat{\delta}(\vec{k},\eta)}{\partial \eta} + \hat{\theta}(\vec{k},\eta) &=& - \int d^3 k_{1} \int d^3 k_2 \delta_{\D}(\vec{k}_{12} - \vec{k}) \alpha(\vec{k}_1, \vec{k}_2) \hat{\theta} (\vec{k}_1, \eta) \hat{\delta} (\vec{k}_2, \eta) \label{massFT} \, , \\
\fr{\partial \hat{\theta}(\vec{k},\eta)}{\partial \eta} + {\cal H} \hat{\theta}(\vec{k},\eta) + \fr{3}{2} \Omega_{m}(\eta) {\cal H}^2 \hat{\delta}(\vec{k},\eta) &=& - \fr{1}{2} \int d^3 k_{1} \int d^3 k_2 \delta_{\D}(\vec{k}_{12} - \vec{k}) \beta(\vec{k}_1, \vec{k}_2) \hat{\theta} (\vec{k}_1, \eta) \hat{\theta} (\vec{k}_2, \eta) \label{EulerFT} \, ,
\ea
where $\vec{k}_{12} \equiv \vec{k}_1 + \vec{k}_2$, $\delta_{\D}$ is the Dirac delta function,  $\Omega_{m}(\eta)$ is the matter energy density contrast, $\alpha(\vec{k}_1, \vec{k}_2) \equiv \fr{\vec{k}_{12} \cdot \vec{k}_1}{k_1^2}$, and $\beta(\vec{k}_1, \vec{k}_2) \equiv \fr{k_{12}^2 (\vec{k}_1 \cdot \vec{k}_2)}{k_1^2 k_2^2}$. Due to the mode coupling of the nonlinear terms, one needs to make a perturbative expansion in $\hat{\delta}$ and $\hat{\theta}$ using the perturbative series of solutions for the fastest growing mode 
\ba \hat{\delta}(\vec{k},\eta) &\equiv& \sum_{n=1}^{\infty} \hat{\delta}^{(n)} (\vec{k},\eta) = \sum_{n=1}^{\infty} D_{1}^{n} (\eta) \delta_{n}(\vec{k},\eta) \nonumber \\ &=& D_{1}^n \int d^3 k_1 \cdots d^3 k_n \delta_{\rm{D}}(\vec{k}_{1 \cdots n} - \vec{k}) C (k_{1}, \cdots, \vec{k}_{n}, \eta) \delta_{1}(\vec{k}_1) \cdots \delta_{1}(\vec{k}_n) \label{hatdeltaS} \, , \\
\hat{\theta}(\vec{k},\eta) &\equiv& \sum_{n=1}^{\infty} \hat{\theta}^{(n)} (\vec{k},\eta) = \sum_{n=1}^{\infty} D_{1}^{n-1} (\eta) D_{1}^{'}(\eta) \theta_{n}(\vec{k},\eta) \nonumber \\ &=& -D_{1}^{n-1} D_{1}^{'} \int d^3 k_1 \cdots d^3 k_n \delta_{\rm{D}}(\vec{k}_{1 \cdots n} - \vec{k}) G_{n}^{(s)} (k_{1}, \cdots, \vec{k}_{n}, \eta) \delta_{1}(\vec{k}_1) \cdots \delta_{1}(\vec{k}_n) \label{hatthetaS} \, , \ea
where the exact solutions for $\delta_{n}$ and $\theta_{n}$ of general dark energy models  are given in Ref. \cite{14077325}. One needs to use the model dependent time varying kernels ($F_{n}^{(s)}$,$G_{n}^{(s)}$) in order to obtain fully consistent results. From above equations (\ref{hatdeltaS}) and (\ref{hatthetaS}), one obtains the one-loop power spectra $P_{\delta \delta}$ and $P_{\delta \delta'}$ 
\ba P_{\delta \delta}(k,\eta) &=& D_{1}^2(\eta) P_{11}(k) + 2 D_{1}^4(\eta) \int d^3 q P_{11}(q) \Biggl[ P_{11}(|\vec{k}-\vec{q}|) \Bigl[ F_{2}^{(s)}(\vec{q}, \vec{k}-\vec{q}, \eta) \Bigr]^2 \nonumber \\ &&+ 3 P_{11}(k) F_{3}^{(s)} (\vec{q},-\vec{q},\vec{k},\eta) \Biggr] \label{Pdeltadelta} \, , \\
P_{\delta \delta'}(k,\eta) &=&  - D_{1}(\eta) D_{1}^{'}(\eta)  P_{11}(k) - D_{1}^3(\eta) D_{1}^{'}(\eta)  \Biggl[ P_{11}(k) \int d^3 q P_{11}(q) \Bigl[ 3 F_{3}^{(s)} (\vec{q},-\vec{q},\vec{k},\eta) + 3 G_{3}^{(s)}(\vec{q},-\vec{q},\vec{k},\eta) \Bigr]  \nonumber \\ 
&&+ 2 \int d^3 q P_{11}(q) P_{11}(|\vec{k}-\vec{q}|) F_{2}^{(s)} (\vec{q},\vec{k}-\vec{q},\eta) G_{2}^{(s)} (\vec{q},\vec{k}-\vec{q},\eta) \nonumber \\ 
&& + 2 \int d^3 q \Bigl[ F_{2}^{(s)} (\vec{q},\vec{k}-\vec{q},\eta) P_{11}(q) P_{11}(|\vec{k}-\vec{q}|)  + G_{2}^{(s)} (\vec{k},\vec{k}-\vec{q},\eta) P_{11}(k) P_{11}(|\vec{k}-\vec{q}|) \nonumber \\ &&+ F_{2}^{(s)} (\vec{k},-\vec{q},\eta) P_{11}(q)P_{11}(k) \Bigr] \alpha(\vec{q},\vec{k}-\vec{q}) \Biggr]\label{Pdotdeltadelta} \ea

\begin{figure}
\centering
\vspace{1.5cm}
\begin{tabular}{cc}
\epsfig{file=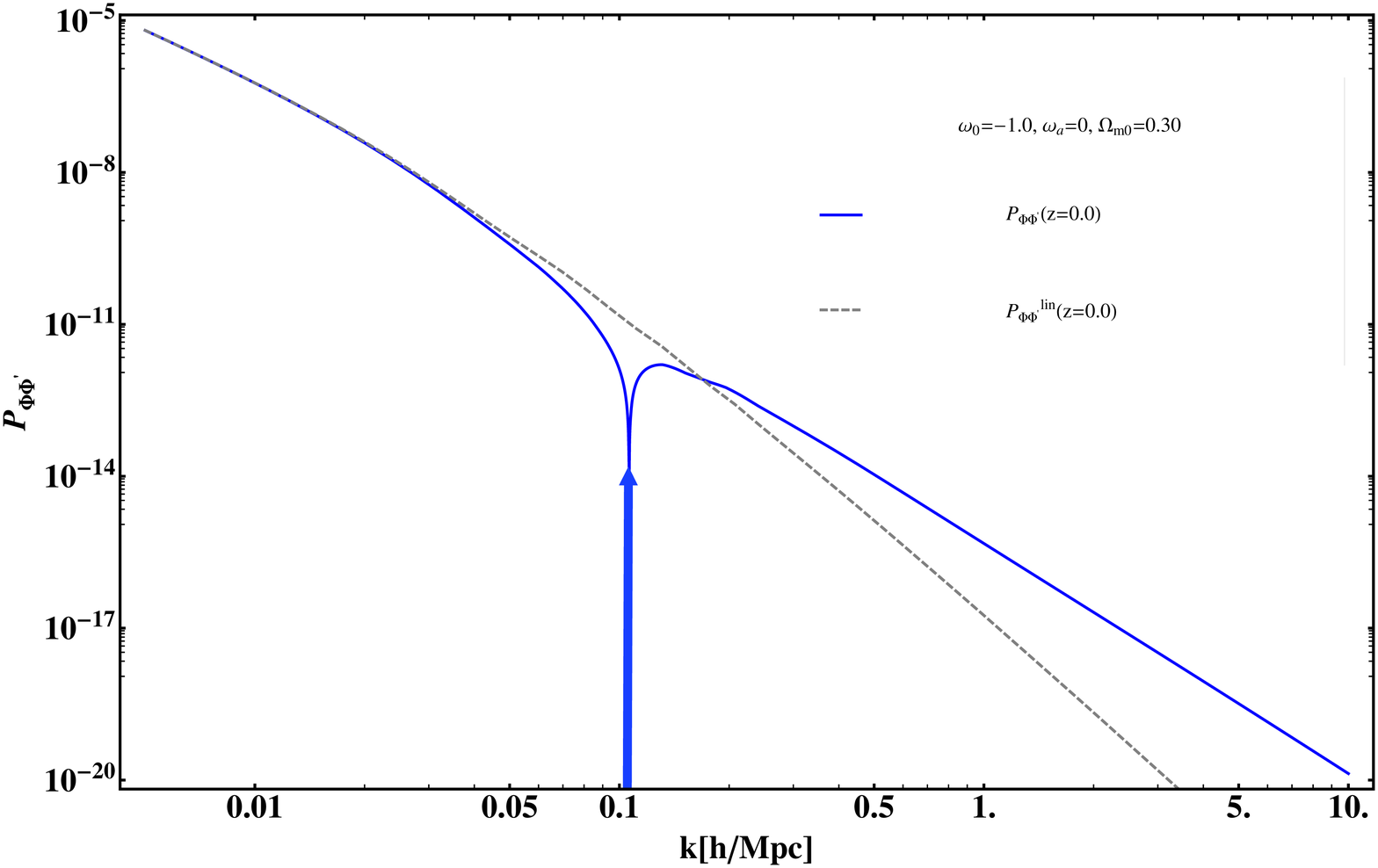,width=0.53\linewidth,clip=} &
\epsfig{file=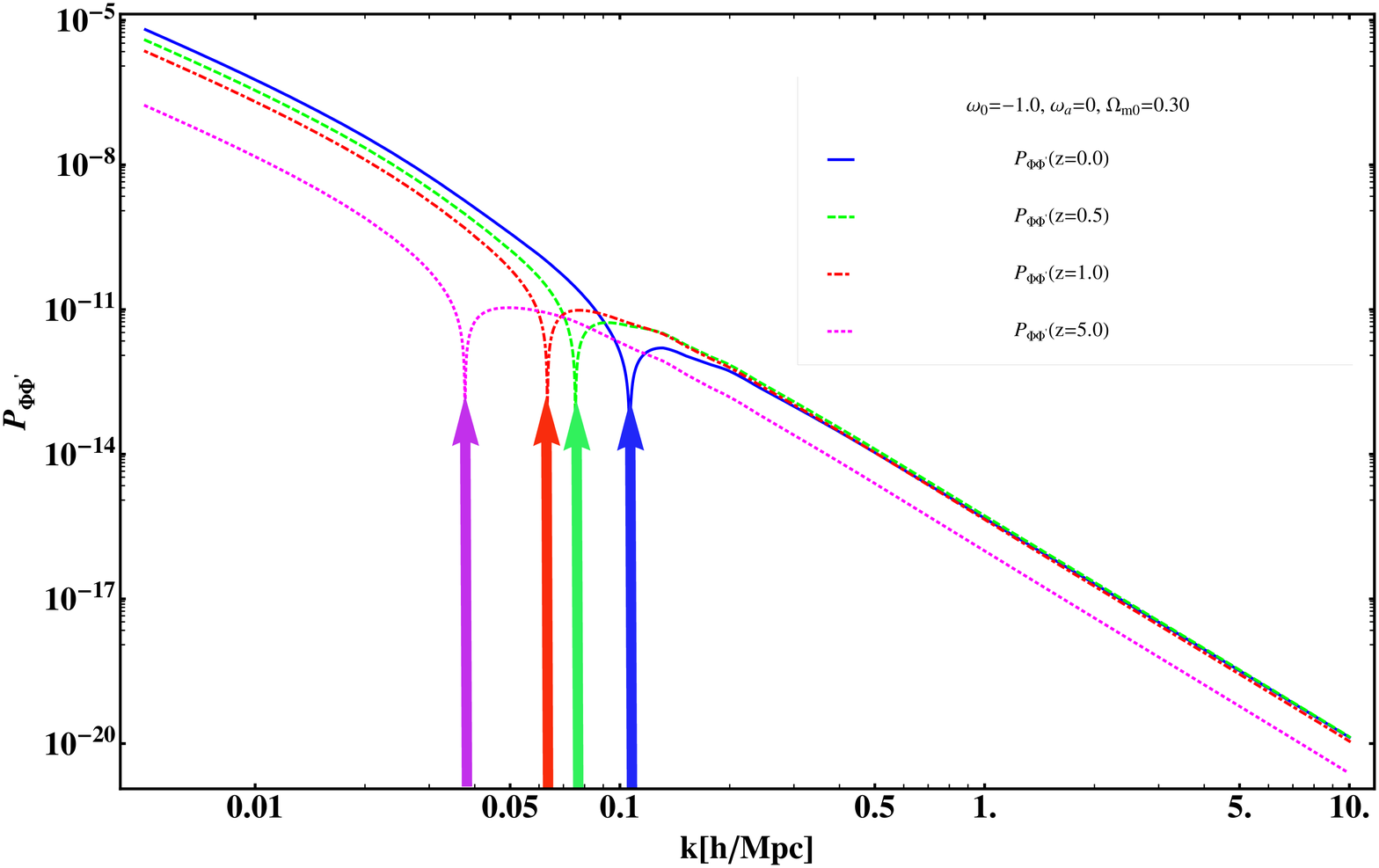,width=0.53\linewidth,clip=} \\
\end{tabular}
\vspace{-0.5cm}
\caption{RS effect for $\Lambda$CDM model with $\Omega_{\rm{m}0} = 0.3$. a) Comparison between ISW effect (dahsed line) and RS effect (solid line) at $z=0$. b) Different flipping scales from anti-correlated to correlated power spectra at different redshifts. The solid, dashed, dotdashed, and dotted lines correspond to $z = 0$, 0.5, 1, and 5, respectively.} \label{fig1}
\end{figure}
One can understand the anti-correlation of linear order $P_{\Phi \Phi'}$ from the above equations (\ref{Pdeltadelta}) and (\ref{Pdotdeltadelta})
\be P_{\Phi \Phi'}^{\rm{Iin}}(k,\eta)  = - \fr{9}{4} \Biggl( \fr{\Omega_{{\rm m} 0}}{a} \Biggr)^2 \Biggl(\fr{H_0}{c k} \Biggr)^4 \Bigl[ D_1 D_1^{'} + {\cal H} D_1^2 \Bigr] P_{11}(k) \label{PPhiISW} \, . \ee All the quantities except sign in Eq. (\ref{PPhiISW}) are positive. Thus, linear order cross-correlation between the ISW effect and the matter density is anti-correlated. From the higher order correction of $P_{\delta \delta'}$, $P_{\Phi \Phi'}$ can be converted into the positively correlated power spectrum from anti-correlated one. This fact is shown in the left panel of Fig.\ref{fig1}. The dashed (solid) line represents the linear (non-linear) power spectrum of $P_{\Phi \Phi'}$. The flip from the anti-correlation to the positive correlation happens around $k = 0.106$ (h/Mpc) at $z =0$ for the $\Lambda$CDM model with $\Omega_{\rm{m}0} = 0.3$. The flipping scales depend on the redshift and cosmological models ($\omega$ and $\Omega_{\rm{m}}$). We show this in the right panel of Fig.\ref{fig1}. The flipping scales are $k = 0.106, 0.076, 0.063$, and 0.038 (h/Mpc) for $z =$ 0 (solid), 0.5 (dashed), 1.0 (dotdashed), and 5 (dashed), respectively.  We will explain these dependences below.  This result is consistent with that of previous works \cite{07111696, 08094488, 14045102}. However, we explicitly obtain the dark energy dependence on the flipping scales which is not obtained in previous works.    

\begin{figure}
\centering
\vspace{1.5cm}
\begin{tabular}{cc}
\epsfig{file=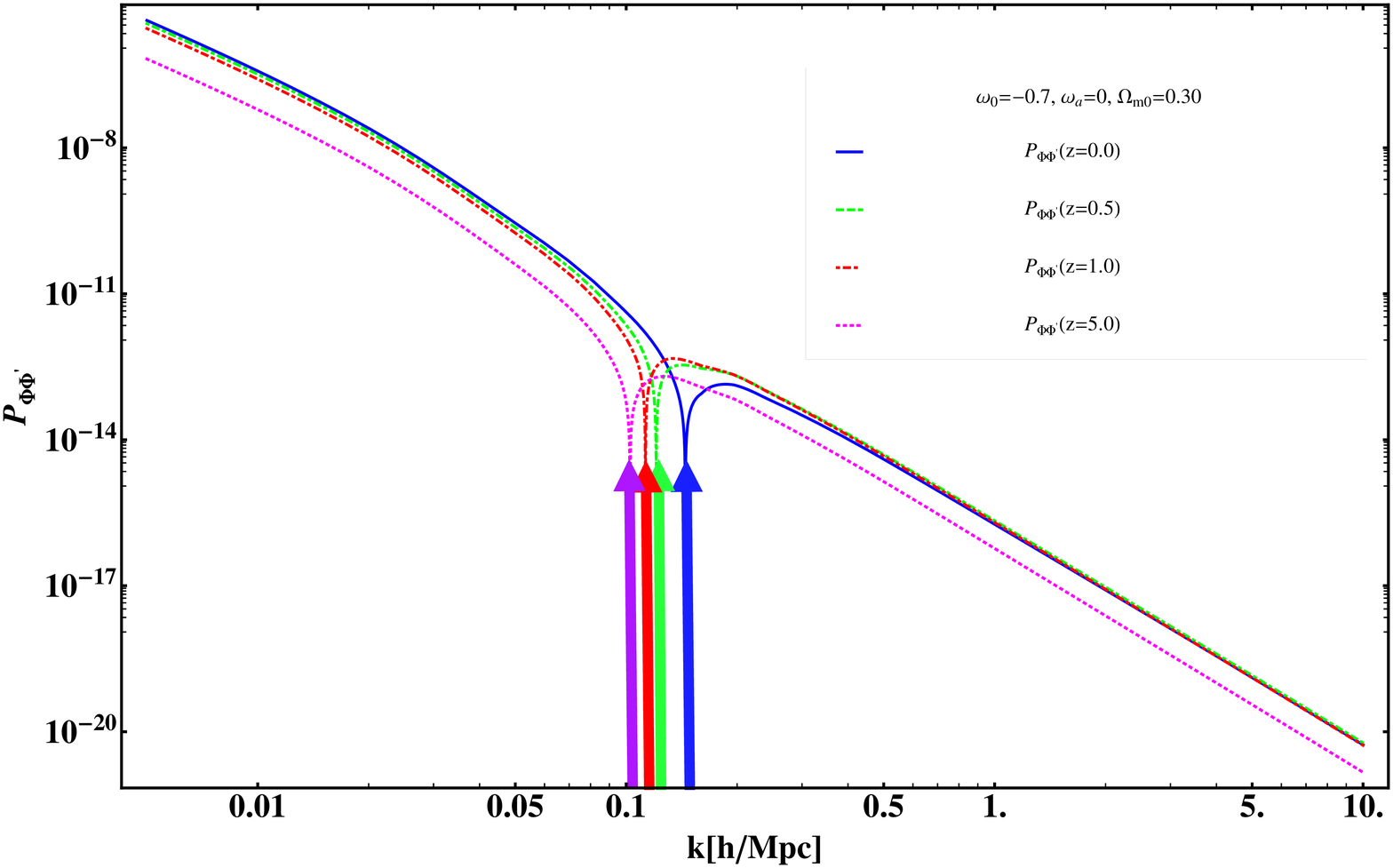,width=0.53\linewidth,clip=} &
\epsfig{file=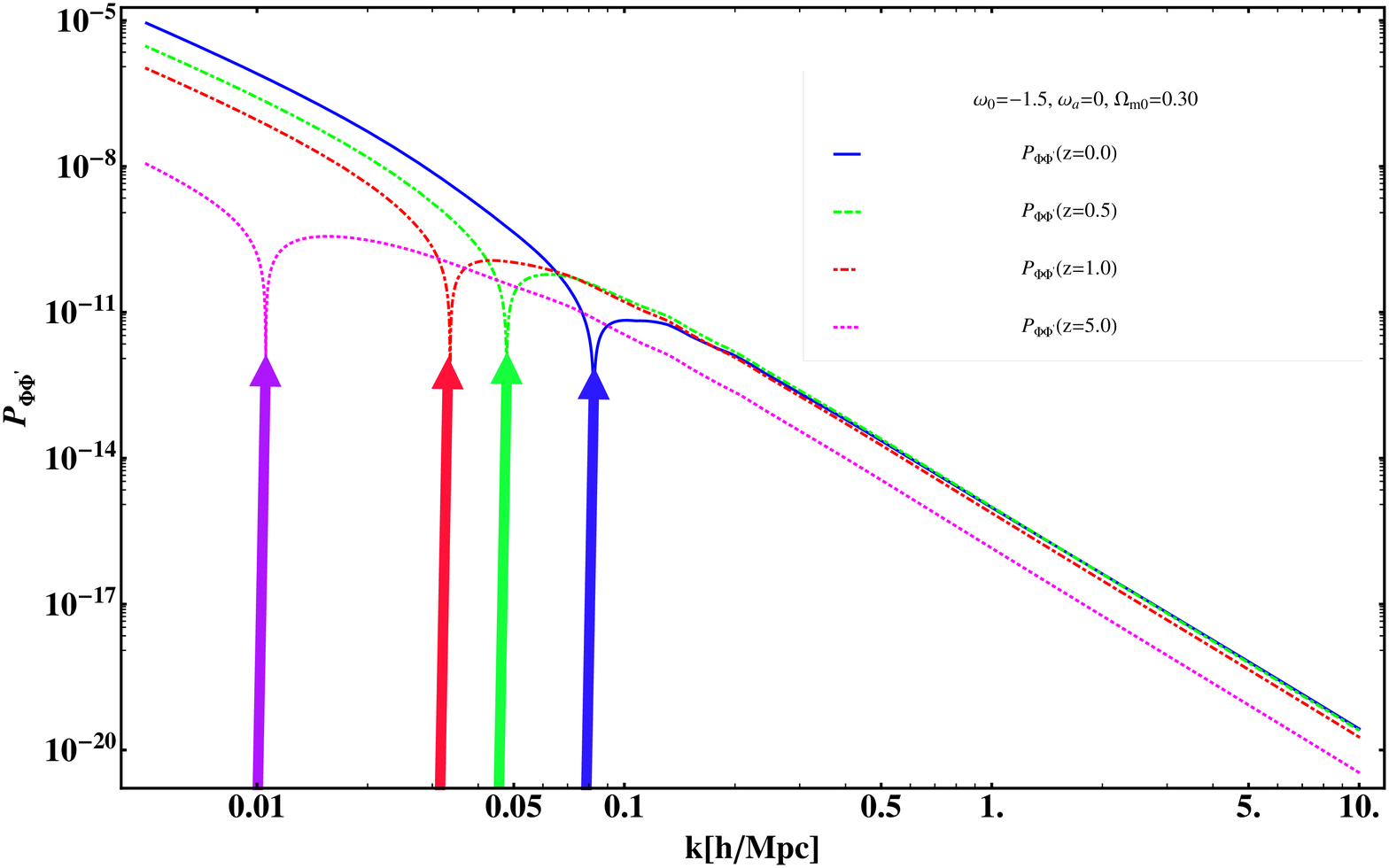,width=0.53\linewidth,clip=} \\
\end{tabular}
\vspace{-0.5cm}
\caption{$P_{\Phi \Phi'}$ showing RS effect for different dark energy models.  a) $P_{\Phi\Phi'}$ for $\omega = -0.7$ at different redshits. The solid, dashed, dotdashed, and dotted lines correspond to $z = 0$, 0.5, 1, and 5, respectively. b) $P_{\Phi\Phi'}$ for $\omega = -1.5$.} \label{fig2}
\end{figure}
We show the dark energy dependence on the flipping scale in the Fig.\ref{fig2}. In the left panel of Fig.\ref{fig2}, we show the power spectrum of $P_{\Phi \Phi'}$ for $\omega = -0.7$ model with $\Omega_{\rm{m}0} = 0.3$. The flipping scales are $k = 0.145, 0.121, 0.113$, and 0.102 (h/Mpc) for $z =$ 0 (solid), 0.5 (dashed), 1.0 (dotdashed), and 5 (dashed), respectively. Compared to $\Lambda$CDM model, flipping scales are smaller for all redshifts. Also intervals between flipping scales in this model are narrower than those of the $\Lambda$CDM. In the right panel of Fig.\ref{fig2}, we show the power spectrum of $P_{\Phi \Phi'}$ for $\omega = -1.5$ model. The flipping scales are $k = 0.083, 0.048, 0.034$, and 0.011 (h/Mpc) for $z =$ 0 (solid), 0.5 (dashed), 1.0 (dotdashed), and 5 (dashed), respectively. Compared to $\Lambda$CDM model, flipping scales are greater for all redshifts. Also intervals between flipping scales are wider than those of the $\Lambda$CDM model.

\begin{figure}
\centering
\vspace{1.5cm}
\begin{tabular}{cc}
\epsfig{file=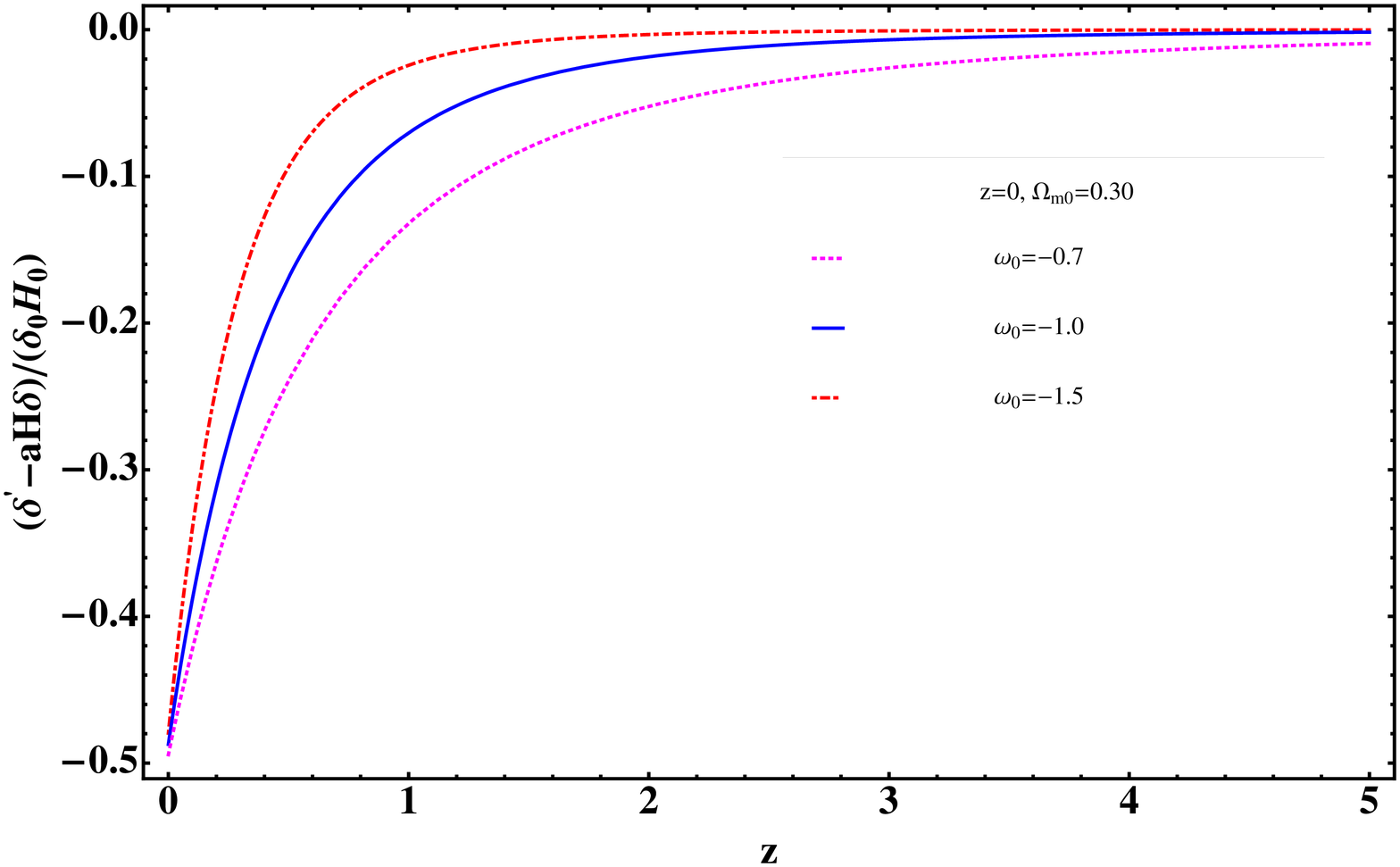,width=0.51\linewidth,clip=} &
\epsfig{file=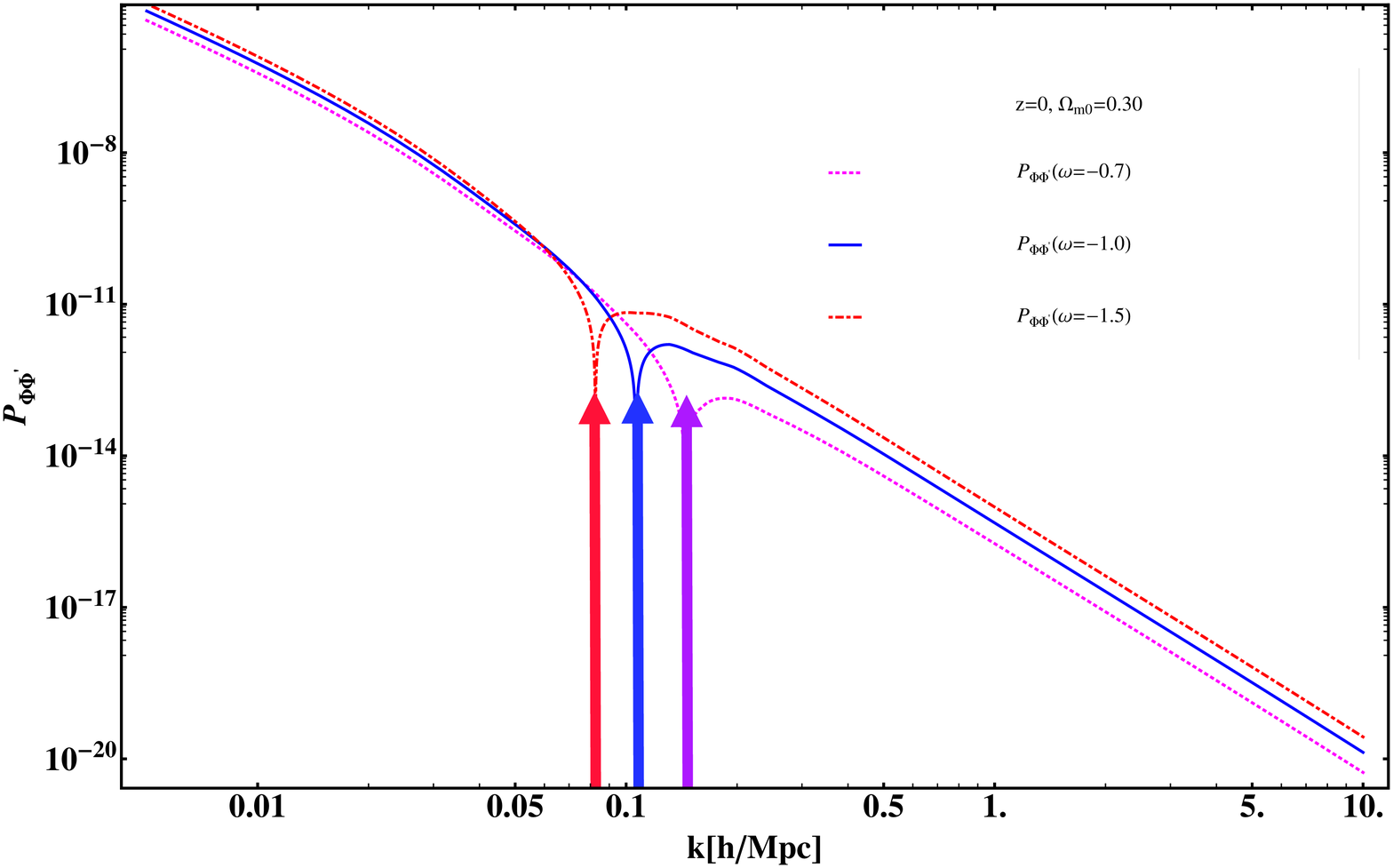,width=0.53\linewidth,clip=} \\
\end{tabular}
\vspace{-0.5cm}
\caption{Model dependence to explain the DE dependence on the flipping scale (see text for details). a)  Model dependence on $(\delta'-{\cal H}\delta)/(\delta H_0)$. The dotted, solid, and dotdashed lines correspond to $\omega = -0.7, -1.0$, and -1.5, respectively a) The flipping scale for the different DE models at z=0 with the same notation as in the left panel.} \label{fig3}
\end{figure}
Now we explain the flipping scales dependence on dark model. As we show, the gravitational potential is always negative for the over density. The linear order time derivative of the gravitational potential is also always positive for the over density region. However, if we include the non-linear contribution of the time derivative of the gravitational potential, then it can be changed to be negative. Thus, the smaller the linear order of $\Phi'$, the larger the non-linear contribution is required to change the sign of $\Phi'$. This is shown in the left panel of Fig.\ref{fig3}. We shows the time evolution of linear gravitational potential $\Phi_{\rm{lin}}^{'} \propto \delta' - {\cal H} \delta$ in this figure. The dotdashed, solid, and dashed lines correspond to $\omega =$ -1.5, -1.0, and -0.7, respectively. The larger the value of $\omega$, the smaller the value of $\delta' - {\cal H} \delta$. Thus, it requires the more non-linear contribution for the larger value of $\omega$ to obtain the sign change of $\Phi'$. Thus, the sign change of the cross power spectrum happens at the smaller scale for the larger value of $\omega$ ({\it i.e.} at the larger $k$). This fact is shown in the right panel of Fig.\ref{fig3}. We use the same notation as the left panel of Fig.\ref{fig3} in this figure. The change of the present epoch power spectrum from the anti-correlation to the positive correlation happens at the largest scale for the smallest value of $\omega$ model. At the present epoch, the flipping scales of the $P_{\Phi \Phi'}$ are 0.083 (for $\omega=-1.5$), 0.106 ($\omega=-1.0$), and 0.145 ($\omega=-0.7$), respectively. 

Also, one understand the dark energy dependence of the width of the intervals of flipping scales at the different redshift from the left panel of Fig.\ref{fig3}.  As shown in the figure, the smaller the value of $\omega$, the steeper the slope of $\delta' - {\cal H} \delta$. Thus, the differences of the required amount of the non-linear contribution at different redshifts becomes larger as the value of $\omega$ decreases. This indicates the wider intervals between flipping scales for the smaller values of $\omega$. These facts are shown through Fig.\ref{fig1} - Fig.\ref{fig3}. Thus, one can use $P_{\Phi \Phi'}$ to investigate the dark energy model. Especially, one can use the location of the flipping scale instead of the amount of $P_{\Phi \Phi'}$ in order to investigate the model dependence. Thus, this can be used as a standard ruler with much less measurement errors. Also, the flipping scales at $z=0.5$ are 0.048, 0.063, and 0.121 h/Mpc for $\omega =$ -1.5, -1.0, and -0.7, respectively. These range of measurements are well measured for the matter power spectrum. One can also predict the flipping scale of any dark energy model at any redshift. Also, one can obtain the flipping scale dependence on $\Omega_{\rm{m}0}$. As $\Omega_{\rm{m}0}$ increases, so does $\delta' - {\cal H} \delta$. Thus, the larger the $\Omega_{\rm{m}0}$, the less non-linear contribution is required. Thus, the flipping scale $k$ will be decreased as $\Omega_{\rm{m}0}$ increases. We summarize these in the table.\ref{Tab1}. 
\begin{table}
\begin{tabular}{|l|c|c|c|c|}\hline
\diagbox[width=2em]{$\omega$}{$z$}& 0 & 0.5 & 1.0 & 5.0 \\ \hline
-0.7 & 0.145 & 0.121 & 0.113 & 0.102 \\ 
-1.0 & 0.106 & 0.076 & 0.063 & 0.038 \\ 
-1.5 & 0.083 & 0.048 & 0.034 & 0.011 \\ \hline
\end{tabular}
\caption{Flipping scales k (h/Mpc) at different redshifts for different dark energy models.}
\label{Tab1}
\end{table}

\section*{Conclusions}
We find the new method for probing the dark energy using Rees-Sciama effect. The cross correlation power spectrum of the ISW effect and mass tracer shows the typical behavior of the change from the anti-correlated to the positively correlated  when it reaches to specific scale. This is due to the fact that the non-linear over density causes the sign change of the time derivative of the gravitational potential. This scale depends on dark energy ($\omega$) and the matter energy density contrast $\Omega_{\rm{m}0}$ at the specific epoch. This method requires the measurement of the location of the sign change of the cross correlation power spectrum instead of the amplitude of it. Thus, it is robust.     

\section*{Acknowledgments}
This work were carried out using computing resources of KIAS Center for Advanced Computation. We also thank for the hospitality at APCTP during the program FRP.

\end{document}